\documentclass[11pt]{article}

\usepackage{graphicx}
\usepackage{bm}
\usepackage{amssymb}
\usepackage{enumerate}

\begin{document}

\def\beq{\begin{equation}}
\def\eeq{\end{equation}}
\def\bea{\begin{eqnarray}}
\def\eea{\end{eqnarray}}
\def\a{\alpha}
\def\b{\beta}
\def\g{\gamma}\def\G{\Gamma}
\def\d{\delta}
\def\e{\epsilon}
\def\phi{\varphi}
\def\k{\kappa}
\def\l{\lambda}
\def\m{\mu}
\def\n{\nu}
\def\o{\omega}
\def\p{\pi}
\def\r{\rho}
\def\s{\sigma}
\def\t{\tau}
\def\th{\theta}
\def\L{{\cal L}}
\def\del{\partial}
\def\nab{\nabla}
\def\half{{\textstyle{\frac{1}{2}}}}
\def\fourth{{\textstyle{\frac{1}{4}}}}

\begin{center} {\LARGE \bf
Hydrodynamics of spacetime \\
and vacuum viscosity}
\end{center}

\vskip 5mm
\begin{center} \large
{Christopher Eling$^{*}$\footnote{E-mail: cteling@phys.huji.ac.il}}\end{center}

\vskip  0.5 cm {\centerline{$^{*}${\it Racah Institute of Physics}}
{\centerline{\it The Hebrew University of Jerusalem}}
{\centerline{{\it Jerusalem 91904, Israel}}}

\vskip 1 cm

\begin{abstract}

It has recently been shown that the Einstein equation can be derived
by demanding a non-equilibrium entropy balance law $dS = \d Q/T +
d_i S$ hold for all local acceleration horizons through each point
in spacetime. The entropy change $dS$ is proportional to the change
in horizon area while $\d Q$ and $T$ are the energy flux across the
horizon and Unruh temperature seen by an accelerating observer just
inside the horizon. The internal entropy production term $d_i S$ is
proportional to the squared shear of the horizon and the ratio of
the proportionality constant to the area entropy density is
$\hbar/4\pi$. Here we will show that this derivation can be
reformulated in the language of hydrodynamics. We postulate that the
vacuum thermal state in the Rindler wedge of spacetime obeys the
holographic principle. Hydrodynamic perturbations of this state
exist and are manifested in the dynamics of a stretched horizon
fluid at the horizon boundary. Using the equations of hydrodynamics
we derive the entropy balance law and show the Einstein equation is
a consequence of vacuum hydrodynamics. This result implies that
$\hbar/4\pi$ is the shear viscosity to entropy density ratio of the
local vacuum thermal state. The value $\hbar/4\pi$ has attracted
much attention as the shear viscosity to entropy density ratio for
all gauge theories with an Einstein gravity dual. It has also been
conjectured as the universal lower bound on the ratio. We argue that
our picture of the vacuum thermal state is consistent with the
physics of the gauge/gravity dualities and then consider possible
applications to open questions.

\end{abstract}

\newpage

\section{Introduction}
\label{intro}

One important clue about the nature of quantum gravity is that the
Einstein field equation and quantum field theory in curved spacetime
imply that black holes behave as thermodynamical objects consistent
with the four laws of thermodynamics. They are endowed with an
entropy proportional to the cross-sectional area of the event
horizon \cite{Bekenstein:1973ur} and a temperature due to quantum
Hawking radiation \cite{Hawking:1974sw}. Although this interplay of
gravitation, quantum field theory, and thermodynamics has been a
focus of research for over 30 years, it has yet to be completely
understood. One fascinating idea was the proposal by Jacobson
\cite{Jacobson:1995ab} to reverse the logic of black hole
thermodynamics and derive the Einstein equations as a consequence of
spacetime thermodynamics and quantum field theory. The idea is that
the local acceleration horizons that exist through every point in
spacetime are analogous to tiny pieces of a black hole event horizon
and have an entropy proportional to their area. The equivalence
principle is then invoked to view the local neighborhood around any
point in a general curved spacetime as a piece of flat spacetime.
Even in flat spacetime accelerated observers can never receive
information from certain regions. For these observers, quantum
fields are localized to the Rindler wedge of flat spacetime and they
view the usual Minkowski vacuum as precisely a thermal state
\cite{Bisognano:1975ih,Unruh:1983ac}. This supplies the notion of a
local temperature. By demanding the Clausius relation $dS = \d Q/T$
at every point, where $\d Q$ is the flow of energy across the
horizon and $T$ is the Unruh temperature, Jacobson was able to show
the Einstein equation appears as an equation of state. This
derivation supports the idea that macroscopic spacetime dynamics is
just the thermodynamics of the quantum vacuum.  In the past several
years other related work has shown that in certain spacetimes
(spherically symmetric, axisymmetric, and cosmological) the
gravitational field equations near a horizon can also be
re-expressed as the thermodynamic identity $T dS = dE + p dV$
\cite{PadFE,Akbar:2006kj}. These results have deep implications, but
their true significance (if any) is not yet clear.

Recently, in order to probe Jacobson's derivation further, a horizon
entropy proportional to a function of the Ricci scalar was
considered \cite{Eling:2006aw}. The idea was to determine whether
higher curvature correction terms that one expects from effective
field theory to appear in the gravitational field equation (or
action) can be derived from the thermodynamical prescription. It was
found that the field equations of $f(R)$ gravity can be derived only
if the setting is shifted from equilibrium to non-equilibrium
thermodynamics\footnote{For another viewpoint where non-equilibrium
thermodynamics is not required in this case, see
\cite{Elizalde:2008pv}}. An extra internal entropy production term
proportional to the squared expansion of the horizon is required for
the Clausius relation (or now more properly the {\it entropy balance
law}) to hold. Perhaps more importantly, it was realized that the
original derivation of the Einstein equation via an area entropy can
be easily generalized to a non-equilibrium setting. In this case an
entropy production term proportional to the squared shear of the
horizon appears when the entropy balance law holds. Consistency with
the Einstein equation requires that the ratio of the shear term
coefficient to area entropy density be $\hbar/4\p$\footnote{We use
units where $c=k_B=1$.}.

The purpose of this paper is to explore the meaning of these
additional terms. The existence of entropy production terms
proportional to squared shear and expansion is reminiscent of the
shear and bulk viscosity terms that appear in viscous fluid
hydrodynamics. A connection between horizon dynamics and fluid
dynamics was first noticed by Damour \cite{Damour}. In the 1980's
Price, Thorne, and collaborators further developed this picture into
the black hole ``membrane paradigm" \cite{membrane}, using a
timelike {\it stretched horizon} to approximate the null horizon.
The shear and expansion of the horizon also appear in the membrane
paradigm, where they are interpreted to be viscous terms in the law
describing how the horizon area/entropy changes. However, this
interpretation is just an analogy and it is not clear if there is a
deep relationship between the dynamics of causal horizons and
hydrodynamics. If there is a relationship we must not only
understand horizons as a fluid system, but also the physics of this
system must be consistent with hydrodynamics as an effective theory.
In hydrodynamics, viscosities are phenomenological coefficients in
the linear constitutive relations between fluxes of momentum in the
fluid and the thermodynamic ``forces" given by gradients of the
fluid velocity. These velocity gradients must be small (both in
space and time) compared to some microscopic scale for hydrodynamics
to be an accurate description of the near-equilibrium physics.

After reviewing the thermodynamic derivation in detail we will show
that it can be consistently reformulated in the language of
hydrodynamics. We argue the Minkowski vacuum, which is a thermal
state when localized in the Rindler wedge of spacetime, is
holographic \cite{'t Hooft:1993gx,Susskind:1994vu}; its properties
are encoded into the 2+1 dimensional system near the Rindler
horizon. This is because the degrees of freedom in this thermal
atmosphere are essentially piled up near the horizon boundary. The
velocity and temperature gradients associated with these degrees of
freedom can be made small enough so that hydrodynamics is the
appropriate effective theory. Therefore we think of these thermal
atmosphere degrees of freedom as a fluid living on a stretched
horizon approximating the local acceleration horizon. Using
hydrodynamics and the properties of the stretched horizon, we will
derive the entropy balance law postulated in \cite{Eling:2006aw},
showing that it is appropriate to interpret the coefficients of the
shear and expansion terms in this law as shear and bulk viscosities
respectively. Just as in the thermodynamic derivation, the entropy
balance law requires that spacetime dynamics is governed by the
Einstein equation and that the shear viscosity to entropy density
ratio is $\hbar/4\p$. Therefore in this more general formulation
Einstein's equation is a consequence of the hydrodynamics of
spacetime vacuum.

The value $\hbar/4\p$ has attracted considerable attention over the
past few years from work on the celebrated anti-de Sitter (AdS)
/conformal field theory (CFT) correspondence \cite{adscft}. This
correspondence states that certain gauge theories in flat spacetime
are equivalent to a quantum gravity theory in a higher dimensional
AdS spacetime. Hydrodynamics arises as the description for the
dynamics of long wavelength perturbations about equilibrium in high
temperature gauge theory plasmas. Working on the gravity side of the
duality it has been found that $\hbar/4\p$ is the value of the shear
viscosity to entropy density ratio for all gauge theories with an
Einstein gravity dual \cite{universality}. Kovtun, Son, and
Starinets conjectured that these theories may saturate a universal
lower bound on the ratio \cite{Kovtun:2004de}, but the true
significance of the ratio is unclear and the existence of a
viscosity bound is controversial \cite{bound}. For instance, it is
not clear why the ratio, derived using relativistic field theories,
is independent of the speed of light $c$. Similarly, if gravity is
somehow involved in saturation of the bound, why does $G_N$ not
appear?

Here, since the ratio holds universally for any local acceleration
horizon, it appears more fundamental than previous results involving
AdS spacetimes. It seems that $\hbar/4\p$ is the shear viscosity to
entropy density ratio of the local vacuum as a thermal state. We
discuss the significance of this result and its connection to the
gauge/gravity literature. We conclude by examining open questions
and possible extensions of our work.

\section{Thermodynamics of spacetime}
\label{thermo}

We now review the derivation~\cite{Jacobson:1995ab, Eling:2006aw} of
the Einstein equation as the equation of state arising from the
thermodynamics of local horizons. The motivating idea is that the
origin of the thermodynamic behavior of black holes is rooted in the
thermodynamic behavior of the local vacuum. In Minkowski spacetime
when quantum fields are restricted to a Rindler wedge $z > |t|$, the
vacuum density matrix takes the form of a thermal state $\rho =
Z^{-1}\exp(-2\p H_B/\hbar)$ \cite{Bisognano:1975ih}. Notice that
neither the ``boost Hamiltonian" $H_B$ nor ``boost temperature"
$T_B=\hbar/2\p$ have dimensions of energy. This is because $H_B$
generates translations of dimensionless hyperbolic boost angle. When
$H_B$ is rescaled to generate proper time along a worldline with
proper acceleration $a$, $T_B$ is rescaled to the usual Unruh
temperature $\hbar a/2\p$ \cite{Unruh:1983ac}.

The null surface $z = t < 0$ forming the edge of the Rindler wedge
is one part of the boundary of the past for the bifurcation plane
$z=t=0$, and therefore acts as a causal horizon. Accelerated
observers in the Rindler wedge can only access information on
spacelike slices bounded by the bifurcation plane. Since vacuum
fluctuations are correlated between the inside and outside of the
wedge, these observers will see an entanglement entropy. In a
continuum field theory this entropy scales with the area of the
boundary, but is divergent because of the ultraviolet (UV)
divergence in the density of states. When a UV regulator is
introduced the entropy is proportional to the area of a horizon
cross-section, with a proportionality constant that could depend on
the nature and number of the quantum fields \cite{entanglement}.
This motivates Jacobson's assumption of a universal entropy density
$s$ per unit horizon area, with $s$ possibly dependent on the field
content. A horizon entropy $\d S= s \d A$ will be contributed by a
little horizon patch of area $\d A$.

When the thermal density matrix $\rho$ at temperature $T_B$ is
perturbed, the change in entanglement entropy is related to the
change in mean energy via
\beq dS = \d <E>/T_B. \eeq
Because the change in the mean energy is due to the flux into the
unobservable region of spacetime, which is perfectly thermalized by
the horizon system, it is assumed to consist entirely of heat. Thus,
we have the thermodynamic Clausius relation $dS = \d Q/T_B$.
Jacobson's second assumption was this relation should hold for all
causal horizons, with $\d Q$ as the flow of boost energy across
the horizon and $dS$ the change in area entropy. Since the area of
the horizon is no longer fixed, the spacetime must become dynamical.

Now that the spacetime is no longer flat everywhere a {\it local}
horizon is defined in analogy with a black hole horizon. A global
definition of the latter is the boundary of the past of future null
infinity. The segment of a black hole horizon to the past of a
spatial cross-section is the boundary of the past of that cross
section. A local horizon at a point $p$ is defined in a similar way:
choose a spacelike 2-surface patch $B$ including $p$, and choose one
side of the boundary of the past of $B$. Near $p$ this boundary is a
congruence of null geodesics orthogonal to $B$. These comprise the
horizon.

At $p$ one can invoke the equivalence principle to view the
spacetime in the neighborhood of $p$ as approximately flat. In this
small patch the idea is to construct a future pointing approximate
boost Killing vector $\chi^\m$ that vanishes at $p$ and whose flow
leaves the tangent plane to $B$ at $p$ invariant. The normalization
of $\chi^\m$ is chosen so that $\chi_{\m;\n} \chi^{\m;\n} = -2$. The
construction can be done explicitly by solving Killing's equation
$\chi_{(\m;\n)} = 0$ order by order in Riemann normal coordinates
$y^\m$. Eventually, at $O(y^3)$, no solution exists because a
general curved spacetime has no Killing vectors. Up to this
ambiguity, our notion of time along the horizon is given by the
parameter $v$ such that $\chi^\m \nab_\m v = 1$. This Killing time
is related to the affine parameter along the horizon generators by
$\l = -e^{-v}$, so the point $p$ is located at infinite Killing time
and $\l=0$. The expansion $\hat{\th}$ and shear $\hat{\s}$ of the
horizon in terms of Killing time are related to the expansion $\th$
and shear $\s$ in affine time as follows
\beq \hat{\th} = e^{-v} \th = -\l \th,~~ \hat{\s}
= e^{-v} \s = -\l \s. \label{falloff}\eeq
Thus, the Killing expansion and shear vanish at $p$ (as long as the
affine quantities are not diverging) and the horizon area is
instantaneously stationary at this point. This defines our notion of
equilibrium.

The system is defined as the degrees of freedom just behind the
horizon and we will consider transitions that terminate in the
equilibrium state at $p$. We define the heat as the flux of the
boost energy current of matter across the horizon,
\beq \d Q = \int T^{M}{}_{\m \n }\chi^\m d\Sigma^\n, \eeq
where $T^{M}{}_{\m \n}$ is the matter stress tensor. This and all
subsequent integrals are taken over a thin pencil of horizon
generators centered on the one that terminates at $p$. It will be
convenient to work in terms of affine parameter and the affinely
parameterized horizon tangent vector $k^\m$. Using the relation
$\chi^\m = - \l k^\m$ and the definition of $T_B$ we thus
have
\beq \frac{\d Q}{T_B} = \frac{2\p}{\hbar}\int T^{M}{}_{\m \n} k^\m
k^\n (-\l) d\l d^2A. \label{dQ/T} \eeq
To compute the entropy change $\d S= s \d A$ we must follow the
area change of the horizon,
\beq \d A = \int \th \, d \l d^2A, \label{dA} \eeq
where $\th=d(\ln d^2A)/d\l$ is the expansion of the
congruence of null geodesics generating the horizon. Using the
Raychaudhuri equation
\beq \frac{d\th}{d\l} = -\frac{1}{2}\th^2-\s_{\m
\n}\s^{\m \n}-R_{\m \n}k^\m k^\n \label{Ray}\eeq
the entropy change is thus given up to $O(\l^2)$ by
\beq \d S = s ~ \int \left[\th
-\l (\frac{1}{2}\th^2+\s_{\m \n}\s^{\m \n}+
R_{\m \n}k^\m k^\n)\right] d\l  d^2A, \label{dS}\eeq
where all quantities in the integrand are evaluated at $p$.

In \cite{Jacobson:1995ab} Jacobson chose $\th = \sigma_{\m \n} = 0$
at $p$, which is required for equilibrium if the affine parameter
$\l$ is assumed to be the natural ``time" of the system. Here and in
\cite{Eling:2006aw} we consider the Killing parameter $v$ to be the
natural time, so vanishing affine expansion and shear at $p$ are not
necessarily needed a priori for our notion of equilibrium. On the
other hand, if it is required that $\d S = \d Q/T_B$ at all points
$p$ and for all null vectors $k^\m$, we first find that the affine
expansion at $p$ must vanish since the heat flux (\ref{dQ/T})
vanishes at $p$. At $O(\l)$ the integrands of (\ref{dQ/T}) and
(\ref{dS}) then imply
\beq (2\p/\hbar) T^{M}{}_{\m \n}k^\m k^\n =  s~(\s_{\m \n} \s^{\m
\n}+ R_{\m \n}k^\m k^\n). \label{Clausiusfinal}\eeq
Note that the shear squared term can be written in terms of derivatives
of $k^\m$, which can be independently chosen at $p$. Therefore the
$k^\m$ derivative part of (\ref{Clausiusfinal}) implies the shear
must also vanish at $p$ if the Clausius relation is to hold.

However, (\ref{falloff}) tells us the shear and expansion with
respect to Killing time fall off to zero at $p$ as $\sim e^{-2v}$
when $\th$ and $\s$ vanish, while only as $e^{-v}$ when $\th$ and
$\s$ are non vanishing. In \cite{Eling:2006aw} it was hypothesized
that for a slower approach to equilibrium the Clausius relation may
not apply and thus $dS > \d Q/T_B$. In this case there is an entropy
balance law
\beq dS = \d Q/T_B+d_i S \label{balance}\eeq
where $d_i S$ represents internal entropy production for a system
out of equilibrium. As is standard in non-equilibrium thermodynamics
we consider the system to be still near enough to equilibrium so
that the entropy and temperature take their local equilibrium
values. The production term $d_i S$ should be of $O(\l)$ to be
consistent with the notion of equilibrium at $p$. We also assume
that $d_i S$ depends only on squared gradients of $k^\m$. Entropy
production from the squared gradients of state variables is a
universal property of non-equilibrium thermodynamics \cite{deGroot}.
Given these assumptions, if the conjectured balance law
(\ref{balance}) is to hold at all $p$ and for all $k^\m$, $\th$ is
still required to vanish and the entropy production term is
\beq d_i S = -s \int \s_{\m \n} \s^{\m \n} \l~ d\l d^2 A. \eeq
In terms of Killing time this new term has the form
\beq d_i S =  s \int \hat{\s}_{\m \n}\hat{\s}^{\m \n} dv d^2 A,
\label{shearprod} \eeq
which looks like the standard \cite{LL} entropy production term for
a fluid with shear viscosity $\eta$,
\beq d_i S = \frac{2\eta}{T_B} \int \hat{\s}_{\m \n}\hat{\s}^{\m \n}
dv d^2 A, \eeq
if we identify $\eta = \hbar s/4\p$.

The remaining $k^\m$ part of the entropy balance law yields
\beq R_{\m \n} + \Phi g_{\m \n} = (2\p/\hbar s)~T^{M}{}_{\m \n}
\label{eos} \eeq
where $\Phi$ is a so far undetermined function. This corresponds to
the tracefree part of the Einstein equation, with Newton's constant
determined by the universal entropy density $s$,
\beq G_N=\frac{1}{4\hbar s}. \eeq
Conversely, $s =1/4\hbar G_N=1/4L_P^2$, so the entropy is identified
as one quarter the area in Planck units, like the Bekenstein-Hawking
black hole entropy. The free function $\Phi$ can be fixed if it is
assumed that the matter stress tensor is divergence free,
corresponding to the usual local conservation of matter energy.
Taking the divergence of both sides of (\ref{eos}) and using the
contracted Bianchi identity $\nab^\n R_{\m
\n}=\frac{1}{2}\nab_\m R$ we then find that $\Phi=-\frac{1}{2}R
- \Lambda$, corresponding to the Einstein equation with
(undetermined) cosmological constant $\Lambda$.

\section{Hydrodynamic Formulation}
\label{hydro}

\subsection{Preliminaries}
\label{prelim}

As noted above, the shear squared entropy production term
(\ref{shearprod}) is very similar to the entropy production term due
to the shear of the fluid flow in viscous hydrodynamics. However, at
this stage it is not at all clear whether it makes sense to identify
$\eta$ as shear viscosity. The shear in (\ref{shearprod}) is a
gradient of a null horizon tangent vector. Can we think of this
horizon tangent as a flow velocity and a cross-section of the
horizon as a ``fluid" system? Also, are we in a near-equilibrium
regime where the horizon shear is small (both in space and time)
compared to some microscopic length scale so that the constitutive
relations are valid and hydrodynamics is an accurate description? It
turns out the answers to these questions are in the affirmative and
it is possible to {\it derive} the entropy balance law and entropy
production term we postulated in (\ref{balance}) and
(\ref{shearprod}) from hydrodynamics.  Before we start, it will be
necessary to briefly review (relativistic) hydrodynamics.

In modern terminology, hydrodynamics is an effective theory
describing the dynamics of perturbations about an equilibrium state
on long length scales. Although the fluid as a whole is not in
equilibrium, we assume it is close enough to it such that
equilibrium states with a local temperature $T$ exist at every
point. ``Long" wavelengths mean long compared to a relevant
microscopic scale for the fluid. This is normally taken as the mean
free path $\ell_{\rm{mfp}}$, which sets the characteristic length
scale over which a system equilibrates locally. In this regime
quantum fluctuations are suppressed and the theory is classical.
Since this is an effective theory our knowledge of the fluid is
restricted to a finite set of variables: typically the equilibrium
proper energy density $\e$ and pressure $P$, temperature $T(x)$ and
local fluid velocity $u^\m(x)$, where $u^\m u_{\m} = -1$. The
hydrodynamic equations are simply the conservation of energy and
momentum $\nab_\m T^{\m \n} = 0$ in the simplest case where there
are no other conserved currents $\nab_{\m} J^\m = 0$ and the
relativistic chemical potential is zero. The stress tensor $T^{\m
\n}$ is a function of the fluid variables and has the form of an
equilibrium perfect fluid plus a dissipative part $\Pi^{\m\n}$
\beq T^{\m \n} = (\e(T)+P(T))g^{\m \n}+P(T) u^\m u^\n + \Pi^{\m \n
}(T, \nab u, \nab T). \eeq
This form of the stress tensor alone is not sufficient to determine
the dynamics. However, if $\nab u$ and $\nab \ln T$ are $\ll
\ell_{\rm{mfp}}^{-1}$, the dissipative part of the stress tensor
$\Pi^{\m\n}$ is smaller than the zeroth order perfect fluid part and
can be expanded in terms of the derivatives of the fluid velocity.
At each point $x$ there is the freedom to boost $u^\m(x)$ such that
$\Pi^{\m \n} u_\n = 0$. With this standard choice of gauge, at first
(linear) order the corrections depend explicitly on the velocity
derivatives and have the form \cite{LL}
\beq \Pi^{\m \n} = P^{\m\a} P^{\n\b} [\eta(\nab_\a u_\b + \nab_\b
u_\a - \frac{2}{3}g_{\a \b} \nab_\g u^\g) + \xi g_{\a \b} \nab_\g
u^\g] \eeq
where $P^{\m \n} = g^{\m \n} + u^\m u^\n$ and we have considered a
fluid of 3 spatial dimensions. In this ``Landau gauge" the spatial
parts of the stress tensor are related to momentum fluxes.
The shear viscosity $\eta$ and bulk viscosity $\xi$ in
this constitutive relation are phenomenological coefficients at this
level and are determined by experiment or matching to the complete
microscopic theory. The derivative expansion can in principle be
continued on to higher orders, with each extra term suppressed by
powers of $\ell_{\rm{mfp}}/L$, where $L$ is the characteristic
length scale of variations in $u^\m$ and $T$.

Returning to the problem at hand, what can we take to be the fluid
system? As stated above, an obvious candidate for the ``fluid" here
is the local acceleration horizon itself, with $k^\m$ as the fluid
velocity. But the technical drawback is the horizon is a null
surface and $k^\m$ is a null vector instead of being unit timelike.
However, we can employ the notion of a stretched horizon pioneered
by Price and Thorne to describe the physics of globally defined
black hole event horizons \cite{membrane}. The stretched horizon is
a timelike surface that lives close enough to the event horizon that
it can capture its essential physics. More specifically, the idea is
to perform a 2+1+1 split of spacetime. The foliating spacelike
surfaces are surfaces of constant time according to a family of
accelerated observers with 4-velocity $U^{\a}$ defined such that
$U_{\m} = \a \ dt$ for lapse function $\a$. In the familiar special
case of a Schwarzschild geometry $\a = (1-2M/r)^{1/2}$ and $t$ are
surfaces of constant Schwarzschild time. The event horizon itself is
characterized by the null generator $l^{\m}$ and can be foliated
into spacelike 2-surfaces by surfaces of constant horizon time. In
the Schwarzschild example $l^{\m} = d/dv$ and $v=\mathrm{const.}$,
where $v$ is the Killing time on the horizon. The distance from the
horizon is naturally parameterized by the affine parameter along the
ingoing null rays (for example just the radial coordinate $r$ in
Schwarzschild), or $\a$ equivalently by a change of variables.

The stretched horizon is defined as a surface of fixed constant
lapse (or radial coordinate) such that $\a \ll 1$. The stretched
horizon itself has unit spacelike normal $N_{\m}$. Since this
vector field can be extended throughout the spacetime as the normal
to all surfaces of constant $\a$, we have a 2+1+1 split defined
by $U^{\m}$ and $N^{\m}$. We will always work in the limit of the
true horizon $\a \rightarrow 0$, where
\bea \a U^{\m} \rightarrow l^{\m} \nonumber\\
\a N^{\m} \rightarrow l^{\m} \label{vectorlimit} \eea
A local stretched horizon in the limit where $\a \rightarrow 0$
will be used to approximate the acceleration horizon, which is the
boundary of the past of $B$. As shown in Figure 1, the stretched
horizon lives just ``inside" the true causal/acceleration horizon.
\begin{figure}
\begin{center}
\includegraphics[angle=270,width=4.4cm]{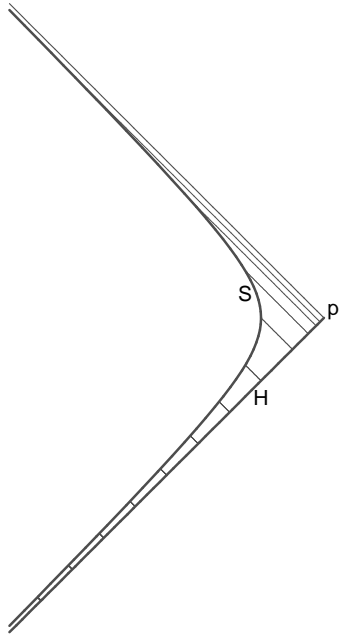}\\
\end{center}
\caption{\label{horizons} \small Representation of how the stretched
horizon $S$ approximates the past horizon $H$. Note that 2 spatial
dimensions are suppressed. Each point in the stretched horizon
represents the fluid at constant time $\t$. As $\a \rightarrow 0$,
points in the stretched horizon are mapped to points in the past
horizon along the ingoing null rays as indicated. A point at
infinite $\t$ is mapped to the bifurcation point $p$.}
\end{figure}
The fluid itself lives on the spacelike cross section of the
stretched horizon defined with the fluid velocity $U^{\m}$}.

Although we have formally identified the stretched horizon system as
a ``fluid", one may wonder if this choice has any physical
interpretation. Since the stretched horizon is a 2+1 dimensional
surface, the fluid must be 2+1 dimensional. In the literature,
\cite{Padsolid} considered the null vector $l^\m$ associated with
null hypersurfaces as an elastic displacement vector of a
``spacetime solid" in a long wavelength limit. The entropy of the
solid is assumed to be a quadratic functional of derivatives of this
vector. When this entropy is maximized gravitational field equations
can be obtained. The idea of a fluid living on a lower dimensional
surface is also consistent with the hydrodynamic limit of AdS/CFT
correspondence. However, there the fluid is taken to be a real gauge
theory fluid living on the timelike boundary of AdS spacetime. Is
there a real fluid on the stretched horizon here?

We argue that the Minkowski vacuum, which looks like a thermal state
when localized into the Rindler wedge of spacetime, obeys the
holographic principle \cite{'t Hooft:1993gx,Susskind:1994vu}. Its
properties are encoded into the 2+1 dimensional stretched horizon
boundary of the wedge. Some similar ideas can be found in
\cite{Padholo}. This appears counterintuitive at first because the
vacuum in the Rindler wedge looks like a 3+1 dimensional bath of
thermal radiation. However, the entanglement entropy associated with
the vacuum is (in the absence of a UV cutoff) a formally divergent
quantity that scales like the cross-sectional area of the horizon
boundary, not the volume of the wedge. Heuristically, the entropy
density of the radiation bath goes like $T^3/\hbar^3$. The key point
is that $T$ is a local Unruh temperature that is a function of the
proper length to the horizon: $\sim \hbar \ell^{-1}$. The total
entropy is
\beq S = \int \ell^{-3} d\ell d^2 A \sim \d A ~ \ell_c^{-2}, \eeq
where $\d A$ is the cross-sectional area of a horizon patch and
$\ell_c$ is the UV cutoff length. A similar calculation for total
energy using density $\sim T^4/\hbar^3$ also yields an area scaling
and stretched horizon energy density $~\hbar \ell_c^{-3}$. Since
these quantities scale like area densities instead of the usual
volume densities they will not be extensive unless we identify them
with the stretched horizon boundary surface. The degrees of freedom
in the vacuum thermal state are effectively packed into the
stretched horizon.

Further evidence for this picture is Brustein and Yarom's work
\cite{BrusteinYarom} showing that vacuum fluctuations in any
sub-volume of Minkowski space scale as the area of the boundary and
diverge unless there is a UV cutoff. They used this result to argue
that these fluctuations have a representation in terms of a high
temperature theory on the boundary, which in the case of the Rindler
wedge, is reminiscent of the near-horizon thermal atmosphere with
its diverging local temperature. In light of the heuristic
arguments above and these results in the literature, we postulate
that if hydrodynamical perturbations of the thermal atmosphere exist
they should be manifested in the dynamics of a stretched horizon. In
particular, in the hydrodynamic limit the degrees of the freedom in
the thermal atmosphere can be represented as a real fluid living on
the boundary.

\subsection{Horizon fluid dynamics}
\label{dynamics}

We now examine the dynamics of the horizon fluid in detail. First we
discuss the fluid in equilibrium. As in Section \ref{thermo} around
the arbitrary point $p$ in $B$ we can invoke the local flat
spacetime approximation to define the local vacuum state. Around
this point we have the approximate set of Poincare symmetries,
including boosts generated by Killing vector $\chi^\a$, which is
defined to vanish at $p$. Using Rindler coordinates adapted to this
boost Killing field $\chi^\m = \frac{\del}{\del \t}$, the
metric in the neighborhood of $p$ has the approximate form
\beq ds^2 \approx - \k^2 \r^2 d \t^2 + d \r^2 + dx^2 + dy^2.
\label{Rindler}\eeq
The lapse function $\a = \k \r$, where $\k$ is an arbitrary constant
associated with the normalization of boost time $\t$. In Section
\ref{thermo}, $\k$ was scaled to be unity and $\t$ was a
dimensionless boost angle. The local Rindler (Killing) horizon at
$\a = 0$ can be approximated by its own stretched horizon for
constant $\a \rightarrow 0$. The Killing vector $\chi^\a$ describes
the horizon fluid rest frame. However there is the freedom to boost
to a moving frame in the $x^i \equiv (x,y)$ directions
\bea \t' &=& \g (\t-\b^i x_i)\\
x'^i &=& \g (x^i - \b^i \t), \label{boost} \eea
to characterize a moving horizon fluid. In this state the flat
spacetime (boosted) Rindler horizon has fixed area and as expected
the entropy is unchanging.

Just as in Section \ref{thermo} we identify the Unruh temperature
$\hbar/2\p \r = \hbar \k/2\p \a$ with the local equilibrium
temperature. Notice this has the Tolman law form $\a T = T_0 =
\mathrm{const.}$, where $T_0 = \hbar \k/2\p$ is analogous to a
position independent Hawking temperature. In this equilibrium state
we expect the fluid is described by a surface stress tensor in the
perfect fluid form
\beq T^{S}_{\m \n} = (\e+P)U_\m U_\n + P \g_{\m \n} \label{stress}
\eeq
where $\g_{\m \n} = g_{\m \n}- N_\m N_\n$ and the superscript $S$
indicates this is a surface tensor. Just like entropy density $s$,
the surface energy density $\epsilon$ and pressure $P$ are formally
divergent quantities that may depend on the number and nature of
fields in the thermal atmosphere. We will allow for a UV cutoff
length $\ell_c$, whose value is initially unknown, which will render
all quantities finite. The stretched horizon boundary metric
((\ref{Rindler}) with $\r$ fixed) is flat and invariant under
translations in time and space. These local translational symmetries
in the boundary imply the surface stress tensor is conserved. Using
the thermodynamic relations $\epsilon+P = sT$, $d\epsilon = T ds$,
and $dP = s dT$ we find the entropy density current
\beq \del_\m(s U^\m) = 0 \label{conserveds}\eeq
is conserved, as expected.

The stretched horizon system of Section \ref{prelim} and the
equilibrium fluid do not agree in general: the fluid velocity $U^\a$
is not proportional to $\chi^\a$ except as $\t \rightarrow \infty$
at $x=y=0$. We have chosen this point because in the limit $\a
\rightarrow 0$, it approaches the bifurcation point at $p$ along the
null ray shown in Figure \ref{horizons}. This supplies the notion of
local equilibrium in the general fluid. For this non-equilibrium
horizon fluid entropy is created externally via heat flux from the
outside of the system, and internally from the friction of
expansions and shears. This implies the horizon area is not fixed,
the entropy current in (\ref{conserveds}) is not conserved, and the
spacetime can no longer be exactly flat. To parameterize the
near-horizon curved metric we follow the construction used by
\cite{Bhattacharyya:2008jc} to study perturbations of black brane
metrics and assume the previously constant $\k$ in (\ref{Rindler})
and boost parameter $\b^i$ in (\ref{boost}) are functions of
stretched horizon coordinates $x^\m \equiv (\t,x,y)$:
\bea \k &\rightarrow& \k(x^\m) \nonumber\\
U^\m &=& \a^{-1}~\g(x^\m) \left(\frac{\del}{\del \t}+ \b^i(x^\m)
\frac{\del}{\del x^i}\right). \eea
$\k(x^\m)$ and the boost parameter $\b^i(x^\m)$ will approach
constant values at $(\infty,0,0)$, where there is no entropy
production and the expansion and shear must vanish.

For hydrodynamics to be an applicable description, the horizon
gradients $\nab_\n \ln \k$ (or equivalently of $\ln T$) and $\nab_\n
\b^i(x^\m)$ (or of $U^\m$) in the local Rindler coordinates need to
be $\ll \ell_{\rm{mfp}}^{-1}$ at $(\infty,0,0)$. By dimensional
analysis the inverse mean free path of this thermal state is
position dependent and $\sim g^2 T/\hbar$, where $g^2$ is an unknown
dimensionless parameter\footnote{In the near horizon limit $\alpha
\rightarrow 0$ the diverging temperature will be much larger than
any other scale.}.

The gradients and the inverse mean free path are divergent as we approach
the true causal horizon $\a \rightarrow 0$, but their ratios are
finite. The horizon gradient of the local temperature is $\nab_\n
\ln T \sim \a^{-1} \nab_\n \ln \k(x^\m)$, while Eqn.
(\ref{vectorlimit}) implies that the gradient has the form
\beq \nab_\n U^\m = \a^{-1} \nab_\n l^\m. \eeq
Thus, we need ${\nab_\n \ln \k, \nab_\n l^\m} \ll g^2 T_0/\hbar \sim
\k g^2$ where now $x^\m = (v,x^i)$ for horizon Killing time $v$.
This criterion is clearly satisfied for derivatives in $v$. This can
be seen because the local equilibration time for the system is $\sim
g^{-2} \k^{-1}$, while the process is assumed to occur for an {\it
infinite} amount of Killing time before terminating in the
equilibrium state. Furthermore in Section \ref{thermo} there was no
requirement on the size of the changes in $x^i$ directions of the
horizon fluid. The stretched horizon cross-section at $\t \rightarrow
\infty$ (or equivalently the 2-surface $B$ as $\a \rightarrow 0$)
can be tuned so that the changes in $\b(x^i)$ and $\ln \k(x^i)$ are
$\ll \k g^2$ near $p$. Thus, there is no obstruction to working in
the hydrodynamic regime and therefore an order by order expansion in
derivatives is justified. In the next subsection we will use the
equations of hydrodynamics and the properties of stretched horizons
to derive the near-equilibrium entropy balance law (\ref{balance})
postulated in Section \ref{thermo}.

\subsection{Entropy balance law and vacuum viscosity}
\label{balancelaw}

Following our above review of hydrodynamics, we can proceed to add a
dissipative part to the perfect fluid stress tensor (\ref{stress})
and expand it in derivatives of the flow velocity. Using
conservation of the stress tensor in the stretched horizon, the
thermodynamic relations $\e+P = sT$, $d \e = T ds$, $dP = s dT$, and
making the gauge choice $\Pi_{\m \n} U^{\n} = 0$, it follows
\cite{LL} that entropy balance law for the horizon fluid is

\beq \del_\m(s U^\m) = \frac{\d Q}{T} + \frac{2\eta}{T}
\tilde{\s}_{\m \n} \tilde{\s}^{\m \n} + \frac{\xi}{T} \tilde{\th}^2, \eeq
where $\tilde{\s}_{\m \n} = \half(\nab_\m U_\n + \nab_\n U_\m
- \th \g_{\m \n})$ and $\tilde{\th} = \nab_\g U^\g$.
The Clausius term is the flux of bulk matter energy into the fluid
as heat. We will see below that the entropy change on the left hand
side of this equation is a finite quantity; the ratios of the
divergent quantities on the right hand side will be finite.

Integrating over a volume in the horizon fluid we find
\bea \int \del_\m (s U^\m) ~\a d\t d^2A = \frac{2\p \a}{\hbar \k}
\int \left[T^{M}{}_{\m \n} U^\m N^\n \a + 2\eta \tilde{\s}_{\m \n} \tilde{\s}^{\m
\n} \a + \xi \tilde{\th}^2 \a \right] d\t d^2A\label{entropybal1}\eea
Using Stokes theorem on the left hand side and then taking the limit
$\a \rightarrow 0$ along with (\ref{vectorlimit}) yields
\beq \d S(v) = \frac{2\p}{\hbar \k} \int \left[T^{M}{}_{\m \n} l^\m
l^\n+ 2\eta \hat{\s}_{\m \n} \hat{\s}^{\m \n} + \xi
\hat{\th}^2\right] dv d^2A, \label{entropybal2}\eeq
where the $\hat{\s}$ and $\hat{\th}$ are now the expansion and shear
of the null $l^\m$. Notice how the $\a$'s have also canceled out of
the right hand side and the relativistic entropy balance law
(\ref{entropybal1}) has been reduced to a non-relativistic form in
the true horizon limit, with the left hand side just a change in
total entropy in Killing time. This result agrees with the equation
for the ``long-time" evolution of black hole entropy in the membrane
paradigm \cite{membrane,Carter}, if we identify $\frac{\hbar \k}{2
\p}$ as a Hawking temperature. What is new here is the conceptual
picture of (\ref{entropybal2}) as a consequence of relativistic
hydrodynamics. This is not present in the Damour-Price-Thorne
membrane paradigm because no hydrodynamic limit was identified
\footnote{In general this limit does not exist. In the case of the
thermal atmosphere outside the horizon of a Schwarzschild black
hole, $\ell^{-1}_{\rm mfp} \sim T_H/\hbar \sim r^{-1}_s$, where
$T_H$ the Hawking temperature. The Schwarzschild radius $r_s$ is the
characteristic size of the system. Spatial gradients of velocity
necessarily scale as $r^{-1}_s$ so there can be no hydrodynamic
limit.}. Thus, $\eta$ and $\xi$ are not just analogous to
viscosities; in our framework it is consistent to identify them as
the shear and bulk viscosity of the horizon fluid.

Working with the bifurcation point parameterized as $v = \infty$ is
not convenient; therefore we change to the affine parameter $\l =
-\k^{-1} e^{-\k v}$ so that $p$ is at the origin: $\l = 0$ and
$x=y=0$. Using the relations $l^a = (d\l/dv) k^a$, $\hat{\th} =
(d\l/dv) \th$, $\hat{\s} = (d\l/dv) \s$, yields
\beq \d S(\l) = -\frac{2\p}{\hbar} \int \left[T^{M}{}_{\m \n} k^\m
k^\n + 2\eta \s_{\m \n} \s^{\m \n} + \xi \th^2\right] \l~ d\l d^2A,
\label{entropybal2}\eeq
which is consistent with the form of the entropy balance law
(\ref{balance}) written in terms of horizon quantities. The matter
stress tensor term is the expected flux of boost energy, while the
viscous terms form the internal production piece $d_i S$. Notice
that in general the horizon bulk viscosity $\xi$ also appears in
addition to the shear viscosity $\eta$.

We can now proceed to expand the left hand side of
(\ref{entropybal2}) order by order in $\l$ just as in the original
thermodynamic derivation discussed at the end of in Section
\ref{thermo}, assuming that the entropy density $s \sim \ell_c^{-2}$
is an undetermined quantity such that $\d S = s \d A$. At zeroth
order the affine expansion at $p$ is again required to be zero.
Demanding the linear order equation hold for all null $k^a$ and at
any arbitrary point $p$ in spacetime, along with local conservation
of the bulk matter tensor $\nab^\n T^{M}{}_{\m \n} = 0$ yields the
Einstein equation and $s$ equal to $(4G_N \hbar)^{-1}$. Thus the UV
cutoff length is fixed to be $\ell_c \sim L_P$. The extra shear term
in the area change that comes from the Raychaudhuri equation
(\ref{Ray}) is consistent with the balance law only if the shear
viscosity to entropy density is $\eta/s = \hbar/4\p$, just like in
the thermodynamical argument. The bulk viscosity $\xi$ is not
determined by the balance law at this order because the affine
expansion must be zero at $p$.

\section{Discussion}

We now conclude with some remarks on the meaning and possible
implications of these results. First, to summarize, we argued that
the thermodynamic derivation of the Einstein field equations can
consistently be reformulated using hydrodynamics. The hydrodynamic
degrees of freedom for the local vacuum state are associated with
local acceleration horizons through any point. These are closely
approximated by surrogate timelike stretched horizons and can be
thought of as 2+1 dimensional fluids. The equivalence principle is
invoked to view the neighborhood of each point as a piece of flat
spacetime with a local Rindler horizon. The entropy of these
horizons is fixed and they describe local equilibrium for the
horizon fluid. The Unruh effect is then used to effectively assign
these equilibrium states a local temperature in the limit where the
stretched horizon approaches the true one. On length scales much
larger than the mean free path, hydrodynamics must be an accurate
description of the physics. When the horizon fluid is out of the
equilibrium state, entropy is produced, the horizon area is no
longer fixed, and the spacetime can no longer be flat. The entropy
balance law is then re-derived using the equations of hydrodynamics.
Together with the local conservation of bulk energy-momentum, the
balance law implies entropy changes must be governed by the Einstein
equation. The Einstein equation thus arises from the hydrodynamics
of the local vacuum. Remarkably, this argument also fixes the
entropy density and shear viscosity of the vacuum such that their
ratio is $\hbar/4\p$.

Our picture seems to imply that microscopic dynamics (which could
include quantum gravity below the cutoff) leads to (semi-)classical
Einstein gravity as collective hydrodynamic behavior at low
energies. Some ideas in the same spirit can be found in
\cite{Volovik}. What is interesting here is that some hydrodynamic
properties turn out to be universal although we initially allowed
for the properties of the horizon fluid to depend on the number and
nature of the quantum fields and treated the viscosities as being
purely phenomenological. Once the value of the the UV cutoff scale
$\ell_c$ was fixed to be roughly a Planck length, the entropy
density associated with all local Rindler horizons is the
Bekenstein-Hawking entropy density and $\eta/s$ is universally
$\hbar/4\pi$. All the dependence on the number and nature of the
quantum fields is apparently absorbed into the low energy Newton
constant $G_N$. This in accord with arguments that the
Bekenstein-Hawking entropy is dependent implicitly on the nature of
quantum fields through the renormalization of the gravitational
constant and is either partly or wholly the entanglement entropy of
the thermal atmosphere \cite{entropy}. These results are puzzling
here since no knowledge of microscopic physics was needed to obtain
them, only the balance law. We did, on the other hand, fix the value
of cutoff scale to be the Planck length ``experimentally" by
requiring the Einstein equation inferred from the entropy balance
law to agree with the observed Einstein equation. Low energy physics
(the balance law) and this one observation turn out to be enough to
determine the entropy density and the shear viscosity of the fluid.
The bulk viscosity though is one fluid property not fixed by the
balance law in this case and therefore it seems one would have to
know about the details of the microscopic physics in order to
determine it.

As we noted in the introduction $\hbar/4\p$ also appears in the
AdS/CFT literature as the universal value of the shear viscosity to
entropy density ratio of gauge theories with an Einstein gravity
dual. Could this be merely a coincidence or is there a connection
between this gauge/gravity duality result and our hydrodynamic
derivation? First, in both cases holography is crucial: we
postulated the thermal vacuum state is holographic, while AdS/CFT is
a precise realization of the equivalence of a higher dimensional
gravity theory to a lower dimensional non-gravitational theory on a
boundary. Furthermore, in the duality, $d$ dimensional gauge
theories in high temperature deconfining phases are dual to large
black hole or black brane spacetimes in $d+1$ AdS
\cite{Witten:1998zw}. Therefore one can use classical perturbations
of the large black hole or black brane spacetimes (see
\cite{Son:2007vk} for a review) to perform analytical computations
of the hydrodynamic transport coefficients. According to the AdS/CFT
dictionary the notion of viscosity is meaningful in the infrared
regime of the gauge theory, which corresponds to the near horizon
limit of the translationally invariant black object. In this sense
these black objects have viscosities, just like the viscosity we
found for local stretched horizons. In both cases the hydrodynamics
of a flat spacetime system is manifested in the dynamics of a
horizon boundary. However, since $\hbar/4\p$ holds for all local
acceleration horizons it seems more fundamental than the AdS/CFT
results for large black holes and black branes in AdS spacetimes.
The dynamics of the local vacuum is governed by gravity itself in
the form of Einstein's equations at each point in an arbitrary
spacetime, while the gauge theory dynamics is encoded in the
perturbation theory about an AdS gravity background.

In the duality the $\hbar/4\p$ result holds for both conformal and
non-conformal gauge theories. The common feature is strong coupling,
in particular very large 't Hooft coupling. In general the shear
viscosity to entropy density ratio of a gauge theory depends on the
value of the 't Hooft coupling \cite{eta/s}. For weakly coupled
theories there is a large separation between the mean free path and
any other microscopic scale, for example a thermal de Broglie
wavelength. In this intermediate region we can use a kinetic theory
description where viscosity is due to momentum transfer by
quasiparticle motion. Larger mean free paths correspond to an easier
momentum transfer and higher viscosity. As the coupling is tuned up
the viscosity decreases and the kinetic theory description begins to
break down. Nevertheless, extrapolating all the way to strong
coupling correctly indicates $\eta/s \sim \hbar$ \cite{Son:2007vk}.

This viewpoint suggests the dynamics of the local vacuum thermal
state should also be strongly coupled in some sense. In fact, using
the results of Section \ref{balancelaw} we can argue this is the
case. From $\eta/s = \hbar/4\p$, we can roughly determine the other
undetermined parameter in our analysis: the dimensionless parameter
$g$ in the mean free path.  First, from kinetic theory we can
estimate $\eta/s \sim \e/s ~l_{\rm{mfp}}$. Using $\e \sim \hbar
\ell_c^{-3}$ and $s \sim \ell_c^{-2}$ for the thermal state, this
implies that
\beq \frac{\eta}{s} \sim \hbar \frac{\ell_{\rm{mfp}}}{\ell_c}.
\label{coupling} \eeq
Thus, consistency with the entropy balance law implies the mean free
path $\ell_{\rm{mfp}}= \frac{\hbar}{g^2 T}$ is of order the UV
cutoff scale $\ell_c$. Since $\ell_c \sim \hbar/T_c$, where $T_c$ is
the local temperature at the cutoff, we find that $g$ must be
roughly of order unity.

We can think of the dimensionless parameter $g^2$ in $\eta/s \sim
\hbar/g^2$  as a coupling which controls the size of the mean free
path compared to the microscopic scale (here the UV cutoff scale).
If $g^2 \ll 1$ the mean free path would be much larger than the UV
cutoff length and the ratio much larger than $\hbar$. However the
hydrodynamic derivation requires $g \sim 1$ and $\ell_{\rm{mfp}}
\sim \ell_c$, which is indicative of strong coupling.

Typically one would not consider the local vacuum thermal state a
strongly coupled system. For example, the vacuum fluctuations of a
free field do not appear to be strongly coupled. On the other hand,
a free (scalar, for example) field theory in flat spacetime is a
continuum field theory and has infinite entanglement entropy. This
conflicts with the requirement of a finite entropy and a non-zero
cutoff that allowed us to derive the entropy balance law\footnote{I
thank Ted Jacobson for suggesting this argument.}. The balance law
implies we must have backreaction effects that distort the flat
background spacetime. These gravitational dynamics are imposed up to
the UV cutoff, where the physics is strongly coupled. In this regime
the entropy density and shear viscosity are universal constants
proportional to one another. Since the bulk viscosity is not
determined, it may depend on the field content. If this is the case
it would be similar to a non-conformal gauge theory, where the bulk
viscosity is determined by the mass scales associated with the
particular fields that break the conformal symmetry.

Since our picture of the local vacuum thermal state and its dynamics
is consistent with key aspects of the gauge/gravity dualities,
perhaps it can provide a new perspective on the puzzling aspects of
the $\hbar/4\p$ ratio. For example, although the ratio is derived
for relativistic field theories, it is independent of the speed of
light $c$ (when we return to cgs dimensions). The hydrodynamic
derivation indicates the ratio is tied crucially to the physics of
null horizons. As we noticed in (\ref{entropybal2}) the entropy
balance law for the local acceleration horizon reduces to a
non-relativistic form. The same type of behavior was first noticed
in the membrane paradigm \cite{membrane}, where the behavior of a
black hole event horizon is analogous to non-relativistic fluid
dynamics. Intuitively, the value of the speed of light should not
affect the behavior of the intrinsically ultra-relativistic degrees
of freedom living on the stretched horizon boundary surface.

An important open question is whether $\hbar/4\p$ is a universal
bound on the shear viscosity to entropy density ratio for all
systems, even those that are non-relativistic. Experimental data and
the fact that viscosity is larger than $\sim \hbar$ for weakly
coupled systems indicate the bound is plausible
\cite{Kovtun:2004de}. It is curious that the conjectured bound is
independent of $G_N$ even though it is saturated in the special
class of theories with a gravity dual. However, in the hydrodynamic
derivation we relied only the general thermal properties of the
Minkowski vacuum. $G_N$ only appears (in the Planck length) when we
require agreement with the experiment and fix $\ell_c \sim L_P$.
This indicates that the bound, if it exists, may be a consequence of
the behavior of quantum fields when they are localized into regions
of flat spacetime. The conjectured bound may also be related to the
Bekenstein entropy bound \cite{Bekenstein:1980jp}, a result also
first derived in a gravitational setting, yet which ultimately does
not depend on $G_N$.

In the future it would be interesting to consider higher
curvature corrections to the assumed entropy density and compare the
results for the viscosities to the gauge/gravity literature. For
example, in the case where the acceleration horizon entropy density
is assumed to be a (non-constant, polynomial) function of the Ricci
scalar $f(R)$ it was found previously that $\eta/s = \hbar/4\p$,
while the bulk viscosity ratio is $3\hbar f/4\p$
\cite{Eling:2006aw}. Corresponding results in the duality would
require study of corrections to non-conformal gauge theories,
although an exact comparison is complicated by our inability to
determine the bulk viscosity even when an area entropy is assumed.
In AdS/CFT, general corrections to Einstein gravity involving
contractions of the Ricci and Riemann tensors were found to modify
the shear viscosity to entropy density ratio at strong 't Hooft
coupling. With a choice of one parameter the ratio can now be less
than $\hbar/4\p$ \cite{Brigante:2007nu,Kats:2007mq}. A comparison
here would require assuming the local horizon has an entropy density
proportional to these contractions and checking the effects in the
entropy balance law (\ref{entropybal2}).

Finally, we have only considered linearized hydrodynamics, which was
sufficient to derive the Einstein equations and fix the shear
viscosity of the local Rindler stretched horizon. However,
\cite{Bhattacharyya:2008jc,Baier:2007ix} recently showed the form of
the hydrodynamic stress tensor on the AdS boundary is determined up
to 2nd order in derivatives by demanding the Einstein equations
(with negative cosmological constant) hold for perturbations about
black brane spacetimes. The procedure is roughly the inverse of our
hydrodynamic derivation: instead of starting with the hydrodynamics
of local Rindler stretched horizon and deriving Einstein's
equations, they impose Einstein's equation order by order in a
derivative expansion to derive the stress-tensor at the AdS
boundary. The resulting perturbative metrics are dual to solutions
of the Navier-Stokes equations. The set of hydrodynamic coefficients
at the next order in the stress tensor characterize relaxation times
\cite{israelstew,causalhydro}. In our case it would be interesting to see
whether higher derivative terms in the stress tensor of the local
Rindler stretched horizon can be meaningfully defined and fixed by
the entropy balance law.

\section*{Acknowledgements}

I am grateful to Jacob Bekenstein and Ted Jacobson for helpful
discussions on drafts of this paper. I also thank Itzhak Fouxon for
a suggestion. This research was supported by the Lady Davis
Foundation at Hebrew University, and by grant 694/04 of the Israel
Science Foundation, established by the Israel Academy of Sciences
and Humanities.

\end{document}